\documentclass[12pt]{iopart}
\usepackage{hyperref}
\usepackage{graphicx}
\usepackage{float} 
\expandafter\let\csname equation*\endcsname\relax
\expandafter\let\csname endequation*\endcsname\relax
\usepackage{amsmath}
\usepackage{color}
\begin{document}

\title[Electrostatic potential of a disk]{Electrostatic potential of a uniformly charged disk  through Green's theorem}

\author{Alina E. Sagaydak$^1$ and Zurab K. Silagadze$^{1,2}$\footnote{Corresponding author.}}
\address{$^1$ Novosibirsk State University, 630 090, Novosibirsk, Russia}
\address{$^2$  Budker Institute of Nuclear Physics, 630 090, Novosibirsk,
Russia}
\ead{a.sagaidak@g.nsu.ru}
\ead{silagadze@inp.nsk.su}

\begin{abstract}
All existing derivations of the electrostatic potential of a uniformly charged disk are technically rather involved.  In an old and now almost forgotten publication, Duffin and McWhirter proposed a method for calculating the electrostatic potentials of planar bodies by a skillful application of Green's theorem. It is shown that this method significantly simplifies the problem of calculating the electrostatic potential of a uniformly charged disk and makes it an almost trivial task.

\end{abstract}

\section{Introduction}
Calculating the electrostatic potential of a uniformly charged disk (or its gravitational analogue) is a classic problem, and there are a variety of ways to solve this problem \cite{Cayley_1874,Durand_1953,Duboshin_1961,Kondratyev_2003,Kondratyev_2007,Krogh_1982,Lass_1983,Conway_2000,Bochko_2020,Ciftja_2011}. However, all existing solution methods are quite complex from a technical point of view, which perhaps explains why this interesting electrostatic problem is not reflected in introductory textbooks on classical electrodynamics.

When the observation point is in the plane of the disk, in \cite{Kondratyev_2003,Kondratyev_2007} the electrostatic potential was calculated by skillful use of Green's theorem \cite{Riley_2006,Stewart_2016}
\begin{equation}
\iint\limits_D\left (\frac{\partial Q}{\partial x}-\frac{\partial P}{\partial y}\right)dxdy=
\oint\limits_{\partial D}\left (Pdx+Qdy\right),
\label{eq1}
\end{equation}
where $P$ and $Q$ must have continuous partial derivatives on an open region that contains $D$.
For our purposes it is sufficient to assume that the plane region $D$ is of the type I, which means that it lies between the graphs of two continuous functions of $x$ \cite{Stewart_2016}. In the line integral, the boundary $\partial D$, assumed to be piecewise-smooth, simple closed curve in the plane, is positively oriented, that is, it is traversed counterclockwise. 

This neat trick replaces the double integral over the surface of the disk with a line integral over the boundary of the disk and really makes calculating the electrostatic potential much easier. However, it cannot be extended in the form in which it is used in \cite{Kondratyev_2007} to observation points located outside the plane of the disk.

\begin{figure}[H]
    \centering
    \includegraphics[scale=0.6]{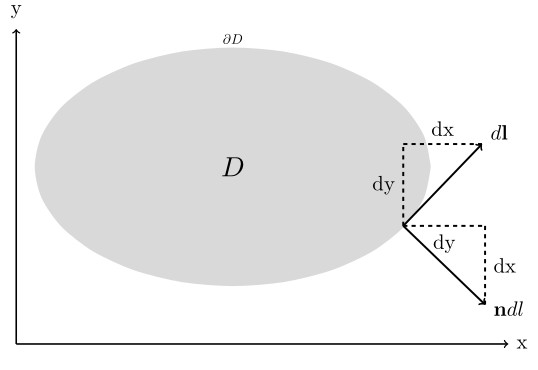}
    \caption{A planar region $D$ bounded by a closed curve $\partial D$, the tangent vector $d{\bf{l}}$ and the outer normal vector ${\bf{n}}\,dl$ at a given point.}
    \label{fig1}
\end{figure}

It is well known that Green's theorem (\ref{eq1}) is equivalent to the two-dimensional divergence theorem \cite{Riley_2006,Stewart_2016}. Indeed, it is clear from Fig.\ref{fig1} that the outer normal ${\bf n}\,dl=dy{\bf i}-dx{\bf j}$, and if we choose ${\bf A}=(Q,-P)$ it is easy to see that
the two-dimensional divergence theorem
\begin{equation}
\iint\limits_D\left (\frac{\partial A_x}{\partial x}+\frac{\partial A_y}{\partial y}\right)dxdy=
\oint\limits_{\partial D}{\bf{A}}\cdot {\bf{n}}\,dl.
\label{eq2}    
\end{equation}
is  transformed into (\ref{eq1}).
Although (\ref{eq1}) and (\ref{eq2}) are equivalent, Duffin and McWhirter \cite{Duffin_1983} showed that the divergence theorem (\ref{eq2}) is more useful and suggestive when used in electrostatic problems for planar charged bodies (they attribute the origin of the idea to Collie \cite{Collie:1976}). Crucial observation of \cite{Duffin_1983} is that, if we take a two-variable scalar function ($z$ considered as a parameter) 
\begin{equation}
\psi(x,y)=r-|z|\ln{(r+|z|)},\;\;\; r=\sqrt{(x-X)^2+(y-Y)^2+z^2},
\label{eq3}
\end{equation}
then
\begin{equation}
\frac{\partial^2 \psi}{\partial x^2}+ \frac{\partial^2 \psi}{\partial y^2}=\frac{r(r+|z|)-(x-X)^2}{r(r+|z|)^2}+\frac{r(r+|z|)-(y-Y)^2}{r(r+|z|)^2}=\frac{1}{r}.
\label{eq4}
\end{equation}
Therefore, taking
\begin{equation}
{\bf A}={\bf \nabla}\psi=\left(\frac{\partial\psi}{\partial x},\frac{\partial\psi}{\partial y}\right ),
\label{eq5}
\end{equation}
in (\ref{eq2}), we find that the electrostatic potential created by a uniformly charged planar region $D$ with a surface charge density $\sigma$ at the observation point $(X,Y,z)$ is given by
\begin{equation} 
\phi(X,Y,z)=\frac{\sigma}{4\pi\epsilon_0}\iint\limits_D\frac{dxdy}{r}=\frac{\sigma}{4\pi\epsilon_0}\oint\limits_{\partial D}\nabla\psi\cdot{\bf n}\,dl. 
\label{eq6}
\end{equation}
Note that the absolute value for $z$ in (\ref{eq3}) is needed to avoid logarithmic singularity when $z<0,x=X,y=Y$. Using (\ref{eq5}), (\ref{eq6}) can also be rewritten as follows 
\begin{equation}
\phi(X,Y,z)=\frac{\sigma}{4\pi\epsilon_0}\oint \frac{(x -X)dy-(y-Y)dx}{\sqrt{(x -X)^2+(y-Y)^2+z^2}+|z|}.
\label{eq6A}
\end{equation}

\section{Calculation of the electrostatic potential}
\begin{figure}[H]
    \centering
    \includegraphics[scale=0.6]{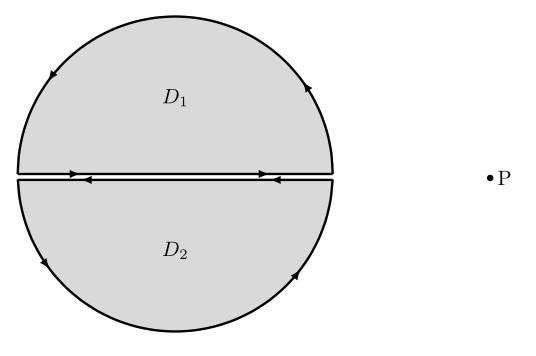}
    \caption{If $P$ is projection of the observation point to the disk plane, $D_1$ and $D_2$ subdomains of the disk contribute equally to the electrostatic potential.}
    \label{fig2}
\end{figure}

Let point $P$ in Fig.\ref{fig2} be the projection onto the disk plane of the point at which we want to find the electrostatic potential (observation point). The coordinate axes can be selected so that $P=(X,Y)=(\eta R,0)$, $R$ being the disk radius and $\eta\ge 0$. In view of the axial symmetry of the problem, to characterize the integration point on the disk, we choose polar coordinates $x=\rho\cos{\alpha},\,y=\rho\sin{\alpha}$. Then $r=\sqrt{(\eta R)^2-2\eta R\rho\cos{\alpha}+\rho^2+z^2}$ and
\begin{equation}
\left .{\bf n}\cdot{\bf\nabla}\psi\right |_{\rho=R}=\left .\frac{\partial \psi}{\partial \rho}\right |_{\rho=R}=\frac{1-\eta\cos{\alpha}}{\sqrt{1+\eta^2-2\eta\cos{\alpha}+z^2/R^2}+|z|/R}.
\label{eq7}
\end{equation}
From the symmetry it is clear that the $D_1$ and $D_2$ halves of the disk make equal contributions to the potential (see Fig.\ref{fig2}). In a line integral, the linear segments of $D_1$ and $D_2$ go in opposite directions, so their contributions cancel each other (alternatively, the symmetry argument can be applied to the line integral itself, stating that the two halves of the circle contribute equally). Therefore, since $dl=Rd\alpha$,
\begin{equation}
\phi(\eta R,z)=\frac{\sigma R}{2\pi\epsilon_0}\int\limits_0^\pi\frac{1-\eta\cos{\alpha}}{\sqrt{1+\eta^2-2\eta\cos{\alpha}+z^2/R^2}+|z|/R}d\alpha=\frac{\sigma R}{2\pi\epsilon_0}\left [I_1-\frac{|z|}{R}\,I_2\right ],
\label{eq8}
\end{equation}
where
\begin{equation}
I_1=\int\limits_0^\pi \frac{(1-\eta\cos{\alpha})\sqrt{1+\eta^2-2\eta\cos{\alpha}+z^2/R^2}}{1+\eta^2-2\eta\cos{\alpha}}d\alpha,
\label{eq9}
\end{equation}
and
\begin{equation}
I_2=\int\limits_0^\pi \frac{1-\eta\cos{\alpha}}{1+\eta^2-2\eta\cos{\alpha}}d\alpha.
\label{eq10}
\end{equation}
The integral $I_2$ can be calculated by a standard substitution $t=\tan{\frac{\alpha}{2}}$ which gives
\begin{equation}
\hspace*{-1mm}
I_2=2\int\limits_0^\infty\frac{1-\eta+(1+\eta)t^2}{(1+t^2)[(1-\eta)^2+(1+\eta)^2t^2]}dt=
\int\limits_0^\infty\left[ \frac{1}{1+t^2}+\frac{1+\eta}{1-\eta}\;\frac{1}{1+\left(\frac{1+\eta}{1-\eta}t\right)^2}\right ] dt.
\label{eq11}
\end{equation}
In the second integral, we make a substitution $s=\frac{1+\eta}{|1-\eta|}t$, where the absolute value is needed to have limits $(0,\infty)$ in $s$. Finally,
\begin{equation}
I_2=\frac{\pi}{2}\left[1+\frac{|1-\eta|}{1-\eta}\right ]=\pi\theta(1-\eta),
\label{eq12}
\end{equation}
where $\theta(x)=\left \{\begin{array}{c} 1, x>0,\\ 0, x<0.\end{array}\right .$ is the Heaviside step function.

Let $\alpha=\pi-2\varphi$. Then (\ref{eq9}) can be rewritten in the following way
\begin{equation}
I_1=2\int\limits_0^{\pi/2}\frac{(1+\eta-2\eta\sin^2{\varphi})\sqrt{(1+\eta)^2+z^2/R^2-4\eta\sin^2{\varphi}}}{ (1+\eta)^2-4\eta\sin^2{\varphi}}\,d\varphi
\label{eq13}
\end{equation}
Denoting temporarily  $A=(1+\eta)^2+z^2/R^2-4\eta\sin^2{\varphi}$, we have
\begin{equation}
1+\eta-2\eta\sin^2{\varphi}=\frac{1}{2}\left [1-\eta^2+A-z^2/R^2\right ],
\label{eq14}
\end{equation}
and
\begin{equation}
I_1=\int\limits_0^{\pi/2}\frac{\left [1-\eta^2+A-z^2/R^2\right ]A}{\left[A-z^2/R^2\right ]\sqrt{A}}\;d\varphi.
\label{eq15}
\end{equation}
But
\begin{equation}
\frac{\left [1-\eta^2+A-z^2/R^2\right ]A}{A-z^2/R^2}=A+(1-\eta^2)\frac{A}{A-z^2/R^2}=A+1-\eta^2+z^2/R^2\frac{1-\eta^2}{A-z^2/R^2}.
\label{eq16}
\end{equation}
Therefore, $I_1=J_1+J_2+J_3$, where 
\begin{equation} 
J_1=\int\limits_0^{\pi/2}\sqrt{A}\,d\varphi=\sqrt{(1+\eta)^2+z^2/R^2}\int\limits_0^{\pi/2}
\sqrt{1-k^2\sin^2{\varphi}}\;d\varphi,
\label{eq17a}
\end{equation}
\begin{equation}
J_2=(1-\eta^2)\int\limits_0^{\pi/2}\frac{d\varphi}{\sqrt{A}}=\frac{1-\eta^2}{\sqrt{(1+\eta)^2+z^2/R^2}}\int\limits_0^{\pi/2}\frac{d\varphi}{\sqrt{1-k^2\sin^2{\varphi}}},
\label{eq17}
\end{equation}
with
\begin{equation}
k^2=\frac{4\eta}{(1+\eta)^2+z^2/R^2},
\label{eq18}
\end{equation}
and
\begin{eqnarray} &&
J_3=\frac{z^2}{R^2}(1-\eta^2)\int\limits_0^{\pi/2}\frac{d\varphi}{\sqrt{A}\,(A-z^2/R^2)}=
\nonumber\\ &&
\frac{1-\eta}{1+\eta}\,\frac{z^2/R^2}{\sqrt{(1+\eta)^2+z^2/R^2}}\int\limits_0^{\pi/2}\frac{d\varphi}{(1-n^2\sin^2{\varphi})\sqrt{1-k^2\sin^2{\varphi}}},
\label{eq19}
\end{eqnarray}
with
\begin{equation}
n^2=\frac{4\eta}{(1+\eta)^2}.
\label{eq20}
\end{equation}
From the definitions of complete elliptic integrals of the first, second and third types \cite{Byrd_2012,Schwalm_2015}
\begin{equation}
K(k)=\int\limits_0^{\pi/2}\frac{d\varphi}{\sqrt{1-k^2\sin^2{\varphi}}},
\label{eq20a}
\end{equation}
\begin{equation}
E(k)=\int\limits_0^{\pi/2}\sqrt{1-k^2\sin^2{\varphi}}\; d\varphi,
\label{eq20b}
\end{equation}
\begin{equation}
\Pi(n,k)=\int\limits_0^{\pi/2}\frac{d\varphi}{(1+n\sin^2{\varphi})\sqrt{1-k^2\sin^2{\varphi}}},
\label{eq20c}
\end{equation}
it is clear that the final expression for the electrostatic potential of a uniformly charged disk, valid for all observation points, takes the form
\begin{equation}
\eqalign{
\;\;\;\;\;\phi(\eta R,z)=\frac{\sigma R}{2\pi\epsilon_0}\left[\frac{1-\eta^2}
{\sqrt{(1+\eta)^2+z^2/R^2}}\,K(k)+\sqrt{(1+\eta)^2+z^2/R^2}\,E(k)+\right .\cr
\left . \;\;\;\;\;
\frac{1-\eta}{1+\eta}\frac{z^2/R^2}{\sqrt{(1+\eta)^2+z^2/R^2}}\,\Pi(n^2,k)-
\pi\,\theta(1-\eta)\,\frac{|z|}{R}\right].}
\label{eq21}
\end{equation}
When calculating the radial component of the electric field from (\ref{eq21}), care must be taken not to get a spurious $\delta$-function singularity, erroneously present in \cite{Bochko_2020}. As explained in \cite{Martin-Luna_2023}, the term containing the elliptic integral of the third kind in the limit $\eta\to 1$ has a discontinuity that compensates for the discontinuity due to the $\theta$-function.

\section{Concluding remarks}
The gravitational analogue of the problem under consideration is of practical interest in astrophysics, since thin disks probably constitute the most numerous class of astrophysical disks. Of course, real astrophysical disks have non-uniform densities. For various reasons such as improper knowledge of boundary conditions, numerical instability of solutions, core singularities, computational cost, etc., the gravitational forces from such disks are not easy to estimate numerically, and this area still presents an interesting challenge \cite{Hure_2011}. However, the analytical results for uniform disks obtained in \cite{Krogh_1982,Lass_1983} have proven very useful in formulating approaches to more realistic situations (see, for example \cite{Tresaco_2011} ). In fact, \cite{Krogh_1982} was initiated by the need to assist in the navigation of Voyager 2 as it passed close to Saturn. ``Without explicit expressions for the gravitational attraction of Saturn's rings, Voyager 2 could not have been so skillfully navigated toward Uranus and then Neptune"  (Krogh F T, personal communication, 2024).

Other real-world applications of the electric potential of a uniformly charged disk include: calculation of the magnetostatic scalar potential and magnetic field of a cylindrical magnet\cite{Martin-Luna_2024}, current distribution on electrode-electrolyte interfaces in electrochemistry \cite{Chen_2020}, molecular beam scattering from a flat liquid jet \cite{Lee_2022}, estimation of the maximum current density extractable from the photocathode \cite{Shamuilov_2018}. Although not directly relevant to the present study, it is interesting to note that Weber's 1873 solution for equipotential disk electric potential is used in analytical studies of axisymmetric, steady-state flows in homogeneous and isotropic aquifers toward a toroidal or disk inlet \cite{Kacimov_2023}.

An interesting question is whether the electric field of charged flat bodies can also be represented as a line integral. Back in 1948, Hubbert showed that this was indeed possible \cite{Hubbert_1948}. Hubbert's method was further elaborated in a seminal paper by Talwani, Worzel and Landisman \cite{Talwani_1959}. At present, this method and a similar method for the magnetic field are perhaps the most widely used methods in geophysics \cite{Blakely_1995}. If a conducting plane body is maintained at a fixed potential in a grounded plane, its electric field is also given by a linear integral of the Biot-Savart law type \cite{Oliveira_2001}. 

If $z=0$, then
\begin{equation}
\frac{\partial}{\partial x}\frac{x-X}{r}+\frac{\partial}{\partial y}\frac{y-Y}{r}=\frac{1}{r},
\label{eq22}
\end{equation}
and we can use Green's theorem (\ref{eq1}) to get
\begin{equation}
\phi(X,Y,z)=\frac{\sigma}{4\pi\epsilon_0}\iint\limits_D\frac{dxdy}{r}=\frac{\sigma}{4\pi\epsilon_0}\oint\limits_{\partial D}
\frac{(x-X)dy-(y-Y)dx}{r}.
\label{eq23}
\end{equation}
Using this formula, Kondratyev in \cite{Kondratyev_2003,Kondratyev_2007} considered a number of electrostatic problems, including uniformly charged circular and elliptical disks. Many other interesting results on potential theory with and without the constraint $z=0$ can also be found in \cite{Kondratyev_2003,Kondratyev_2007}.

The use of Green's theorem for planar electrostatic problems in the form proposed in \cite{Duffin_1983} makes calculating the electrostatic potential of a uniformly charged disk in terms of complete elliptic integrals an almost trivial task. 

The reason for this simplification is that the method fully exploits the cylindrical symmetry of the problem. If this symmetry is broken, such as in the case of an elliptical disk, the calculations will not be as simple.

The electrostatic potential of a uniformly charged elliptical disk was found in \cite{Kondratyev_2003,Kondratyev_2007}. Of course, it will be interesting to reproduce this result using Green's theorem in the form (\ref{eq2}).

A drawback of the method under consideration, as presented in this article, is that its application is limited only to the case of uniform charge density.

Nevertheless, we hope that students will benefit from this approach and that this interesting electrostatic problem will finally find its place in introductory classical electrodynamics textbooks.

\appendix

\section{Electric potential on the axis of symmetry of a regular polygonal plate}
Using Green's theorem for planar electrostatic problems is especially useful when there is symmetry in the problem. To illustrate this and compare with similar calculations \cite{Ciftja_2020,Ciftja_2023,Sheng_2023,Okon_1982,Fagundes_2022,Rao_1979}, let us calculate the electric potential on the axis of a uniformly charged $N$-sided regular polygonal plate.

\begin{figure}[H]
    \centering
    \includegraphics[scale=0.6]{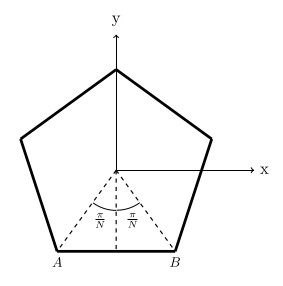}
    \caption{A plate in the shape of a regular polygon.}
    \label{fig3}
\end{figure}

For edge $AB$, $y=-\frac{a}{2}\cot{\frac{\pi}{N}}$, $\vec{n}dl=-dx\vec{j}$, where $\vec{j}$ is the unit vector in the $y$ direction, and $a$ is the length of the side of the polygon (see Fig.\ref{fig3}).
Therefore, the contribution of the edge $AB$ to the electric potential on the $z$ axis is given by 
\begin{equation}
\varphi_{AB}=-\frac{\sigma y}{4\pi\epsilon_0}\int\limits_{-a/2}^{a/2}\frac{dx}{r+|z|}=-\frac{\sigma y}{2\pi\epsilon_0}\int\limits_0^{a/2}\frac{dx}{r+|z|},
\label{A1}
\end{equation}
where $r=\sqrt{x^2+y^2+z^2}$. To calculate the integral, it is useful to use the decomposition \cite{Duffin_1983}
\begin{equation}
\frac{1}{r+|z|}=\frac{r(r-|z|)}{r(r^2-z^2)}=\frac{x^2+y^2+z^2-r|z|}{r(x^2+y^2)}=\frac{1}{r}+\frac{z^2}{r(x^2+y^2)}-\frac{|z|}{x^2+y^2}.
\label{A2}
\end{equation}
The first and the last integrals are immediately calculated:
\begin{equation}
\int\limits_0^{a/2}\frac{dx}{\sqrt{x^2+y^2+z^2}}=\operatorname{arcsinh}{\frac{a}{2\sqrt{y^2+z^2}}},\;\; \int\limits_0^{a/2}\frac{dx}{x^2+y^2}=\frac{1}{y}\arctan{\frac{a}{2y}}.
\label{A3}
\end{equation}
The middle integral is calculated by successive substitutions $\sqrt{x^2+y^2}=|z|\tan{t}$, $\cos^2{t}=\tau$ and $\tau=\frac{z^2-s^2}{z^2+y^2}$. The result is
\begin{equation}
\int\limits_0^{a/2}\frac{dx}{r(x^2+y^2)}=\frac{1}{|z|y}\arctan{\frac{|z|a}{y\sqrt{a^2+4y^2+4z^2}}}.
\label{A4}
\end{equation}
Therefore,
\begin{equation}
\varphi_{AB}=-\frac{\sigma y}{2\pi\epsilon_0}\left [ \operatorname{arcsinh}{\frac{a}{2\sqrt{y^2+z^2}}}+\frac{|z|}{y}\left ( \arctan{\frac{|z|a}{y\sqrt{a^2+4y^2+4z^2}}}-\arctan{\frac{a}{2y}}\right)\right ].
\label{A5}
\end{equation}
Because of symmetry, all edges of a regular polygon contribute equally, and finally the electric potential on the axis of symmetry of a regular polygonal plate takes the form
\begin{equation}
\varphi=\frac{\sigma a N}{4\pi\epsilon_0}\left [\cot{\frac{\pi}{N}} \operatorname{arcsinh}{\frac{1}{\sqrt{\cot^2{\frac{\pi}{N}}+\frac{4z^2}{a^2}}}}+\frac{2|z|}{a}\left(\arctan{\frac{2|z|\tan{\frac{\pi}{N}}\sin{\frac{\pi}{N}}}{\sqrt{a^2+4z^2\sin^2{\frac{\pi}{N}}}}}-\frac{\pi}{N}\right)\right ].
\label{A6}
\end{equation}
If $z=0$, using $\operatorname{arcsinh}{x}=\ln{\left (x+\sqrt{1+x^2}\right )}$, we get
\begin{equation}
\varphi(0,0,0)=\frac{\sigma a N}{4\pi\epsilon_0}\cot{\frac{\pi}{N}}\ln{\left ( \sec{\frac{\pi}{N}}+\tan{\frac{\pi}{N}}\right )},
\label{A7}
\end{equation}
in agreement with \cite{Sheng_2023}.  

If $N=4$, using relations
\begin{equation}\label{A8}
    \begin{split}
        \arctan{A}-\arctan{B}&=\arctan{\frac{A-B}{1+AB}},\;\;AB > -1,\\
    \arctan{A}&=\frac{1}{2}\,\arctan{\frac{2A}{1-A^2}},\;\;A^2 < 1,
    \end{split}
\end{equation}
    
we get
\begin{equation}
\varphi=\frac{\sigma a}{\pi\epsilon_0}\left [ \operatorname{arcsinh}{\frac{a}{\sqrt{a^2+4z^2}}}-\frac{z}{a}\arctan{\frac{a^2}{2z\sqrt{2a^2+4z^2}}}\right ],
\label{A9}
\end{equation}
in agreement with \cite{Ciftja_2023}. 

\section*{References}
\bibliography{charged_disk}

\end{document}